\documentclass[aps, pre, twocolumn, superscriptaddress, nofootinbib]{revtex4}

\usepackage[english]{babel}
\usepackage{hyperref}

\usepackage[pdftex]{graphicx}
\usepackage{color}
\usepackage[toc,page]{appendix}

\usepackage{amsmath}
\usepackage{amssymb}
\usepackage{mathtools}
\usepackage{braket}

\newcommand{\un}[1]{\ensuremath{\,\mathrm{#1}}}
\renewcommand{\v}[1]{\ensuremath{\boldsymbol{#1}}}
\newcommand{\dg}{\dagger}
\newcommand{\fig}[1]{Fig.~\ref{fig:#1}}
\newcommand{\Fig}[1]{Fig.~\ref{fig:#1}}
\newcommand{\sect}[1]{Section~\ref{sec:#1}}
\newcommand{\eq}[1]{(\ref{eq:#1})}
\newcommand{\lr}[1]{\ensuremath{\left( #1 \right)}}

\renewcommand{\Im}[1]{\ensuremath{\mathrm{Im} \left(#1\right)}}
\newcommand{\I}{\mathrm{i}}
\renewcommand{\ap}{\alpha}
\newcommand{\Gm}{\Gamma}

\newcommand{\sg}{\sigma}

\newcommand{\mc}{\mathcal}
\newcommand{\Abs}[1]{\ensuremath{\left| #1 \right|}}
\newcommand{\Tr}[1]{\ensuremath{\mathrm{Tr}\left(#1\right)}}
\newcommand{\pd}{\partial}

\newcommand{\eps}{\varepsilon}
\newcommand{\bt}{\beta}
\newcommand{\dl}{\delta}
\newcommand{\abs}[1]{\left| #1 \right|}

\newcommand{\cT}[1]{\textcolor{black}{#1}}

\begin{document}

\title{Current splitting and valley polarization in elastically deformed graphene}

\author{Thomas Stegmann}
\email{stegmann@icf.unam.mx}
\affiliation{Instituto de Ciencias F\'isicas, Universidad Nacional Aut\'onoma de M\'exico, Cuernavaca, Mexico}

\author{Nikodem Szpak}
\email{nikodem.szpak@uni-due.de}
\affiliation{Fakult\"at f\"ur Physik, Universit\"at Duisburg-Essen, Duisburg, Germany}

\date{\today}

\begin{abstract}
  Elastic deformations of graphene can significantly change the flow paths and valley polarization
  of the electric currents. We investigate these phenomena in graphene nanoribbons with localized
  out-of-plane deformations by means of tight-binding transport calculations. Such deformations can
  split the current into two beams of almost completely valley polarized electrons and give rise to
  a valley voltage. These properties are observed for a fairly wide set of experimentally accessible
  parameters. We propose a valleytronic nanodevice in which a high polarization of the electrons
  comes along with a high transmission making the device very efficient.  In order to gain a better
  understanding of these effects, we also treat the system in the continuum limit in which the
  electronic excitations can be described by the Dirac equation coupled to curvature and a
  pseudo-magnetic field.  Semiclassical trajectories offer then an additional insight into the
  balance of forces acting on the electrons and provide a convenient tool for predicting the
  behavior of the current flow paths.  The proposed device can also be used for a sensitive
  measurement of graphene deformations.
\end{abstract}

\maketitle

\section{Introduction}
\label{sec:Introduction}

One of the remarkable features of graphene and some other 2D materials is the fact that the
electrons occupy the valleys around two inequivalent Dirac points $K^\pm$ in momentum space. This
additional valley spin degree of freedom can be used for a new type of electronics, called
valleytronics \cite{Schaibley2016}. Its key element is an efficient valley polarizer which can
separate in space the electrons from the two different valleys.  Due to the high interest in this
field, several proposals have been already suggested: a graphene valley polarizer can be constructed
by using gated constrictions \cite{Rycerz2007, Jones2017}, the trigonal warping of the Dirac cones
\cite{Garcia-Pomar2008, Yang2017}, a line defect \cite{Gunlycke2011, Ingaramo2016}, electrostatic
potentials in bilayer graphene \cite{Costa2015}, or artificial electron masses with
\cite{Grujic2014} or without \cite{Costa2017} spin-orbit coupling.

Recently, it has been investigated how mechanical deformations affect the electronic properties of
graphene, see \cite{Vozmediano2010, Amorim2016, Naumis2017} for an overview. It has been shown that
a valley polarizer can be constructed by using strained graphene in combination with electrostatic
gates \cite{Jiang2013} or magnetic barriers \cite{Fujita2010, Yesilyurt2016}. Moreover, strain alone
can lead to valley polarization of the electrons. The required strain patterns are generated, for
example, by out-of-plane folds \cite{Carrillo-Bastos2016}, bumps \cite{Settnes2016, Milovanovic2016}
or in-plane wavelike deformations \cite{Cavalcante2016}. Some of these deformations have been
realized experimentally: bumps \cite{Lee2008, Wong2010, Shin2016, Smith2016} and folds
\cite{Huang2011b} have been produced in suspended graphene, a bump-like deformation pattern has been
engineered recently \cite{Nemes-Incze2017, Georgi2017} in graphene on a substrate.

\begin{figure}[t]
  \centering
  \includegraphics[scale=0.36]{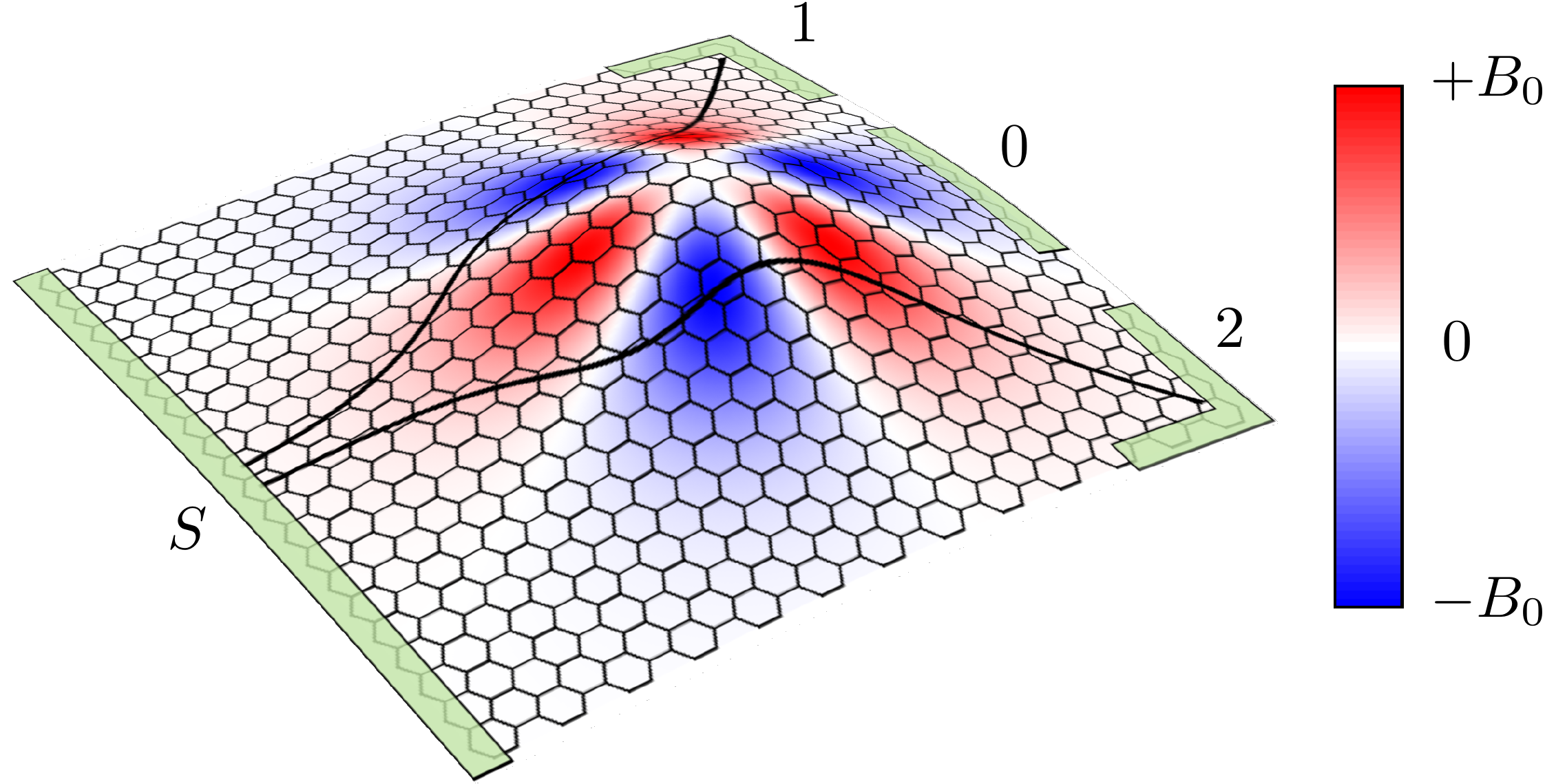}
  \caption{The current flow in a graphene nanoribbon with a localized out-of-plane deformation
    \eq{2} is studied. Electrons are injected through a wide source contact on the left zigzag
    edge. The transmission of these electrons through the system is detected by three contacts
    $\{ 0,1,2\}$ on the right hand side. The deformation creates a pseudo-magnetic field which acts
    with the opposite signs on the two valleys $K^\pm$ and can be used to generate valley-polarized
    currents. The classical trajectories of the electrons in the two different valleys are indicated
    by the black solid lines.  Note that in order to show the orientation of the graphene lattice,
    the system has been scaled down compared to its actual size of $100 \times 100 \un{nm}$.}
  \label{fig:1}
\end{figure}

The essential idea of those approaches is to use the pseudo-magnetic field which is produced by the
strain. The pseudo-magnetic field for a bump-like deformation is sketched in \fig{1} by the red and
blue color shading. It is oriented perpendicular to the graphene sheet and acts with opposite signs
on the two valleys. This valley dependency can be used to separate spatially the electrons from the
different valleys. The curved black trajectories in \fig{1} indicate the deflection of the electrons
injected at the source contact $S$ in different valleys $K^\pm$. The valley polarized electrons can
then be extracted at the contacts $1$ and $2$.

The general idea to construct an all-strain based valley polarizer has been presented recently in
\cite{Settnes2016, Milovanovic2016}. Here, we extend and significantly improve the previous work. We
suggest a new device in which almost complete valley polarization of the electrons comes along with
a high transmission (conductance) making our device very efficient. This is achieved because we do
not filter out from the injected electrons a small part of valley polarized electrons but we split
up all the current into two valley-polarized beams. \cT{In contrast to studies of quasi-infinite
  systems \cite{Chaves2010, Settnes2016}, we attach realistic contacts to the edges of the device in
  order to quantify the current flow in the system and assess its efficiency. In this way, we obtain
  a highly efficient valley polarizing device working in a wide range of experimentally accessible
  parameters.}

In our previous work we have developed a geometric language, based on the continuous elasticity
theory, for describing current flow paths in deformed graphene \cite{Stegmann2016}.  That approach
makes use of the Dirac equation coupled to effective curvature and pseudo-magnetic field.
Semiclassical trajectories provide an efficient method to estimate the current flow paths and will
also be used to gain additional insight into the system.

\section{Device and its modeling}
\label{sec:System}

We consider a graphene nanoribbon of size $\lr{L_x,L_y}= \lr{100 \times 100}\text{nm}$. The
nanoribbon is described by the tight-binding Hamiltonian
\begin{equation}
  \label{eq:1}
  H= -\sum_{n,m} t_{nm} \ket{n^A}\bra{m^B} +\text{h.c.},
\end{equation}
where $\ket{n^{\text{A/B}}}$ indicate the atomic states localized on the carbon atoms at positions
$\v r_n$ on sublattices A or B, respectively. The sum runs over nearest neighboring atoms. In
homogenous graphene these atoms are separated by a distance of $d_0=0.142 \un{nm}$ and coupled with
the energy $t_{nm}= t_0= 2.8 \un{eV}.$

The nanoribbon is deformed elastically by lifting the carbon atoms according to the height function
\begin{equation}
  \label{eq:2}
  h(\v r )= \frac{h_0}{1+\lr{\abs{\v r -\v r_c}/r_0}^2}.
\end{equation}
The center of the deformation is defined by $\v r_c$. Its height and width are controlled by the
parameters $h_0$ and $r_0$, respectively. Experimentally, such deformations can be generated by
substrate engineering and the tip of atomic force microscope \cite{Lee2008, Wong2010,
  Nemes-Incze2017} or gas pressurized membranes \cite{Settnes2015, Milovanovic2017, Shin2016,
  Smith2016}. The modification of the coupling matrix elements is in good approximation described by
\begin{equation}
  \label{eq:3}
  t_{nm} \cong t_0 \exp(-\bt \dl_{nm}),
\end{equation}
where $\dl_{nm}= \frac{\abs{\v r_n- \v r_m}-d_0}{d_0}$ and $\bt=3.37$ \cite{Pereira2009,
  Ribeiro2009, Carrillo-Bastos2016}.

\subsection{The effective Dirac equation in curved space}
\label{sec:EffectiveDirac}

\cT{At low energies, where the electron wavelength is much larger than the lattice constant, and for
  small deformations the discrete tight-binding Hamiltonian \eq{1} can be approximated by the
  continuous Dirac Hamiltonian%
  \footnote{\cT{The spin-connection term, which guarantees hermiticity of the Hamiltonian, can be
      neglected as a higher order correction. The Dirac equation naturally couples to the curved
      geometry via a local frame field \cite{Juan2012, Stegmann2016, Juan2013, Oliva-Leyva2015}. In
      \cite{Stegmann2016}, we demonstrated that the waves propagate along geodesics and therefore,
      the effective geometry can be assumed to be Riemannian with the metric tensor obtained from
      the frame. However, for strong deformations the effective current paths may deviate from
      geodesics and require an extended geometric language, including for example the torsion
      \cite{Zubkov2013, Volovik2015}.}}
\begin{equation} 
  \label{eq:18}
  H^D = \I \hbar v_F \sg^a e_a^{\;\;l}(\v x) \lr{\pd_l - \I K^{\pm}_l(\v x)}.
\end{equation}
describing relativistic massless fermions in curved space \cite{Juan2012, Juan2013, Oliva-Leyva2015,
  Stegmann2016}.} Here, $v_F=3 t_0 d_0/2\hbar$ is the Fermi velocity of the excited electrons and
$\sg^a$ ($a=1,2$) are the Pauli matrices. The local frame vectors $\v e_a(\v x)$ are determined by
the effective strain tensor $\hat\eps = \bt\, \hat\eps_0$, which transforms the local frame
\begin{equation}
  \label{eq:local-frame}
  \v e_a(\v x) = \lr{1 - \bt \hat\eps_0(\v x)} \v e_a.
\end{equation}
Hence, the effective geometry for the electronic excitations is only magnified by the scalar factor
$\bt>1$ but is otherwise identical to the real geometry of the deformed nanoribbon. $\v K^\pm(\v x)$
is a vector potential given by \cite{CastroNeto2009, Vozmediano2010}
\begin{equation}
  \v K^\pm(\v x) = \v K^\pm \pm \frac{\beta}{2} 
  \begin{pmatrix}
    -2 \eps_{xy},& \eps_{yy}-\eps_{xx},
  \end{pmatrix}
\end{equation}
where $\v K^\pm= (0, \pm\frac{4 \pi}{3\sqrt{3} d_0})$ are two Dirac points of pristine graphene. The
curl of this vector potential gives rise to an effective pseudo-magnetic field
\begin{equation}
  B^\pm(\v x) =  \pm [\pd_x\, \eps_{yy}(\v x) - \pd_x\, \eps_{xx}(\v x)  + 2\pd_y\, \eps_{xy}(\v x)],
\end{equation}
which is perpendicular to the graphene plane. In contrast to a true magnetic field, the
pseudo-magnetic field acts with the opposite sign in the two different valleys and hence, the
time-reversal symmetry of the system is preserved. The change of sign of the pseudo-magnetic field
will be used to separate spatially the electrons from the two different valleys, see \fig{1}. The
finite curvature expressed by the local frame $\v e_a(\v x)$ also affects significantly the
transport \cite{Stegmann2016}. However, it acts equivalently on the two valleys and hence has no
effect on the valley polarization.

\subsection{Current flow lines in the geometric optics approximation}
\label{sec:FlowLines}

\cT{At low energies and for large-scale deformations, which fulfill the hierarchy
\begin{equation*}
  \label{eq:geoaprox}
  \text{lattice constant} \ll \text{wavelength} \ll \text{deformation scale},
\end{equation*}
the continuum and geometric optics approximations can be applied to the tight-binding Hamiltonian
\eq{1}.} In our previous work \cite{Stegmann2016}, we have shown that in that case the current flow
in deformed graphene can be predicted by trajectories of relativistic massless fermions which move
in a curved space in the presence of a pseudo-magnetic field
\begin{equation} \label{eq:geodesic}
   \frac{dv^i}{d\tau} = - \Gamma^i_{kl} v^k v^l + \sqrt{g}\ g^{ij} \epsilon_{jk}\, v^k B^{\pm}
\end{equation}
where $v^i(\tau) = dx^i(\tau)/d\tau$ is the ``velocity'' and the right-hand side terms describe
geometric and magnetic forces, respectively ($\Gamma^i_{kl}$ are Christoffel symbols and
$\epsilon_{ij}$ is the Levi--Civita symbol).

The calculation of these trajectories is computationally much less demanding than the quantitative
quantum approach described in \sect{NEGF} and independent from the system size. Therefore, it
provides a useful tool to estimate the current flow in deformed graphene nanostructures, as shown in
\fig{1}.

\subsection{The nonequilibrium Green's function method for the current flow}
\label{sec:NEGF}

The current flow in graphene nanoribbons is studied quantitatively by means of the nonequilibrium
Green's function (NEGF) method. \cT{This quantum method is based on the tight-binding Hamiltonian
  \eq{1} with hopping parameters \eq{3} modified due to the deformation and a model for the contacts
  at which the electrons are injected and detected. It does not rely on the Dirac approximation of
  \sect{EffectiveDirac} nor on the geometric optics approximation of \sect{FlowLines} and hence,
  allows us to verify their validity.} As the NEGF method is discussed in detail in various
textbooks, see e.g. \cite{Datta1997, Datta2005}, we summarize here only briefly the essential
formulas.

The Green's function of the system is given by
\begin{equation}
  \label{eq:4}
  G(E)= \lr{E-H-\Gm}^{-1},
\end{equation}
where $E$ is the single-particle energy of the injected electrons \cT{and $H$ is the tight-binding
  Hamiltonian \eq{1}}. The self-energies $\Gm= \sum_i \Gm_i$ describe the effect of the contacts
attached to the nanoribbon, see the green bars in \fig{1}.

The electrons are injected at the left ribbon edge as plane waves propagating to the right with
momentum $\v q = (q_x,0) = (\abs{E}/\hbar v_F,0 )$.  This injection is modeled by the inscattering
function
\begin{equation}
  \label{eq:6}
  \Gm_S^{\text{in}}= \sum_{n,m \in S} g(\v r_n) g(\v r_m) {\psi^D_{n}}^*(\v q) \psi^D_{m}(\v q) \ket{n} \bra{m},
\end{equation}
where the sum runs over all carbon atoms in contact with the source.  The $\psi_n^D(\v q)$ are the
eigenstates of the Dirac Hamiltonian \eq{18} \cT{(without any deformation)}
\begin{equation}
  \label{eq:7}
  \psi^D_n(\v q){=}
  \begin{cases}
    c_{-} e^{i(\v q{+}\v K^-)\v r_n} + c_{+} e^{i(\v q{+}\v K^+)\v r_n} & n \in A,\\
    \sg c_{-} e^{i(\v q{+}\v K^{-}) \v r_n + i \phi} \,{-}\, \sg c_{+} e^{i(\v q{+}\v K^{+}) \v r_n {-} i\phi} \hspace*{-2mm}& n \in B,
  \end{cases}
\end{equation}
where $\phi= \arg(\I q_x + q_y)$ and $\sg=\text{sign}(E)$. The $c_\pm$ are the amplitudes of the
excitations around the two $\v K^\pm$ valleys and are used to control the valley polarization of the
injected current. The function
\begin{equation}
  \label{eq:7b}
  g(\v r)= e^{-4\log(2)(y-L_y/2)^2/d_{\text{opt}}^2}
\end{equation}
gives the injected current beam a Gaussian profile. The parameter
$d_{\text{opt}}^2= \frac{2 \pi \hbar v_F}{\abs{E}}L_x$, which controls the width of the Gaussian
beam, is chosen in such a way that the beam shows minimal diffraction \cite{Stegmann2016}. \cT{Note
  that using the eigenstates of flat graphene is justified, because $\Gm_{S}^{\text{in}}$ has
  nonzero matrix elements only at the left ribbon edge, where the source contact is located and the
  deformation vanishes.} We assume that the injection of the electrons does not affect their
propagation in the graphene nanoribbon and hence, we take $\Gm_S=0$.

Three contacts are attached at the right ribbon edge, where the electrons are absorbed and the
current flow can be measured, see \fig{1}. For these contacts we use the wide-band model with
\begin{equation}
  \label{eq:5}
  \Gm_j= -\I \sum_{n \in C_j} \ket{n}\bra{n} \quad \text{and} \quad
  \Gm_j^{\text{in}}= -2\, \Im{\Gm_j},
\end{equation}
where the sum runs over the atoms that are connected to the contact $j \in (0,1,2)$. Moreover, in
order to suppress boundary effects, we attach a virtual wide-band contact to those edge atoms which
are not connected to a real contact.

The local current flowing between the carbon atoms is calculated by
\begin{equation}
  \label{eq:9}
  \mc{I}_{nm}= \Im{t_{nm}^* G^{\text{in}}_{nm}},
\end{equation}
where 
\begin{equation}
  \label{eq:10}
  G^{\text{in}}= G\, \Gm_S^{\text{in}}\, G^\dg.
\end{equation}
The transmission (or conductance) between the source and one of the three contacts at the right edge
($i \in (0,1,2)$), is given by
\begin{equation}
  \label{eq:11}
  T_{iS}= \Tr{\Gm_i^{\text{in}}\,G\,\Gm_S^{\text{in}}\,G^\dg}.
\end{equation}
The transmission depends not only on the properties of the nanoribbon but also on the size and the
model of the contacts attached to it. For a better comparison, we normalize in the following the
transmission with respect to the total transmission $T_{\text{tot}}=\sum_{i \in{0,1,2}} T_{iS}$ in a
flat nanoribbon.

\begin{figure}[t]
  \centering
  \includegraphics[scale=0.4]{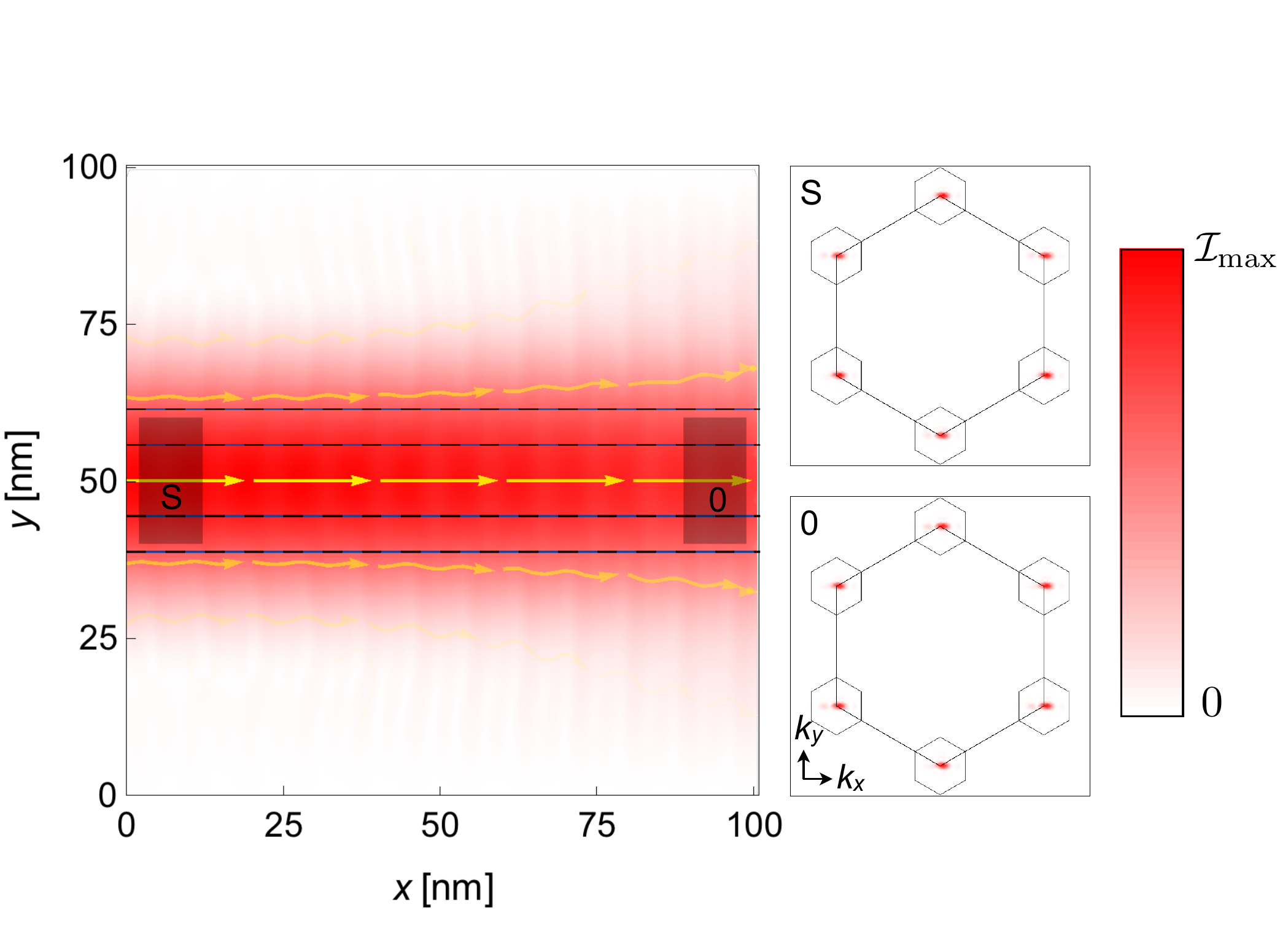}
  \caption{Left: Current flow in a flat graphene nanoribbon, where electrons are injected at the
    left edge with energy $E= 225 \un{meV}$. The current (its density is indicated by the red color
    shading and its vector field by the yellow arrows) goes straight along the system indicating
    ballistic conduction. The current flow lines in geometric optics approximation are indicated by
    the blue-black dashed curves. Right: The projection $\mc{P}_i(\v k)$ in the gray-shaded
    rectangular regions $i \in (S,0)$ shows that the current is composed of states from all six
    valleys and hence, is completely unpolarized. The small hexagons indicate the regions over which
    the spectral density is integrated to measure the polarization. Note also that the regions
    around the valleys have been magnified by the factor $2.5$.}
  \label{fig:2}
\end{figure}

\begin{figure}[t]
  \centering
  \includegraphics[scale=0.4]{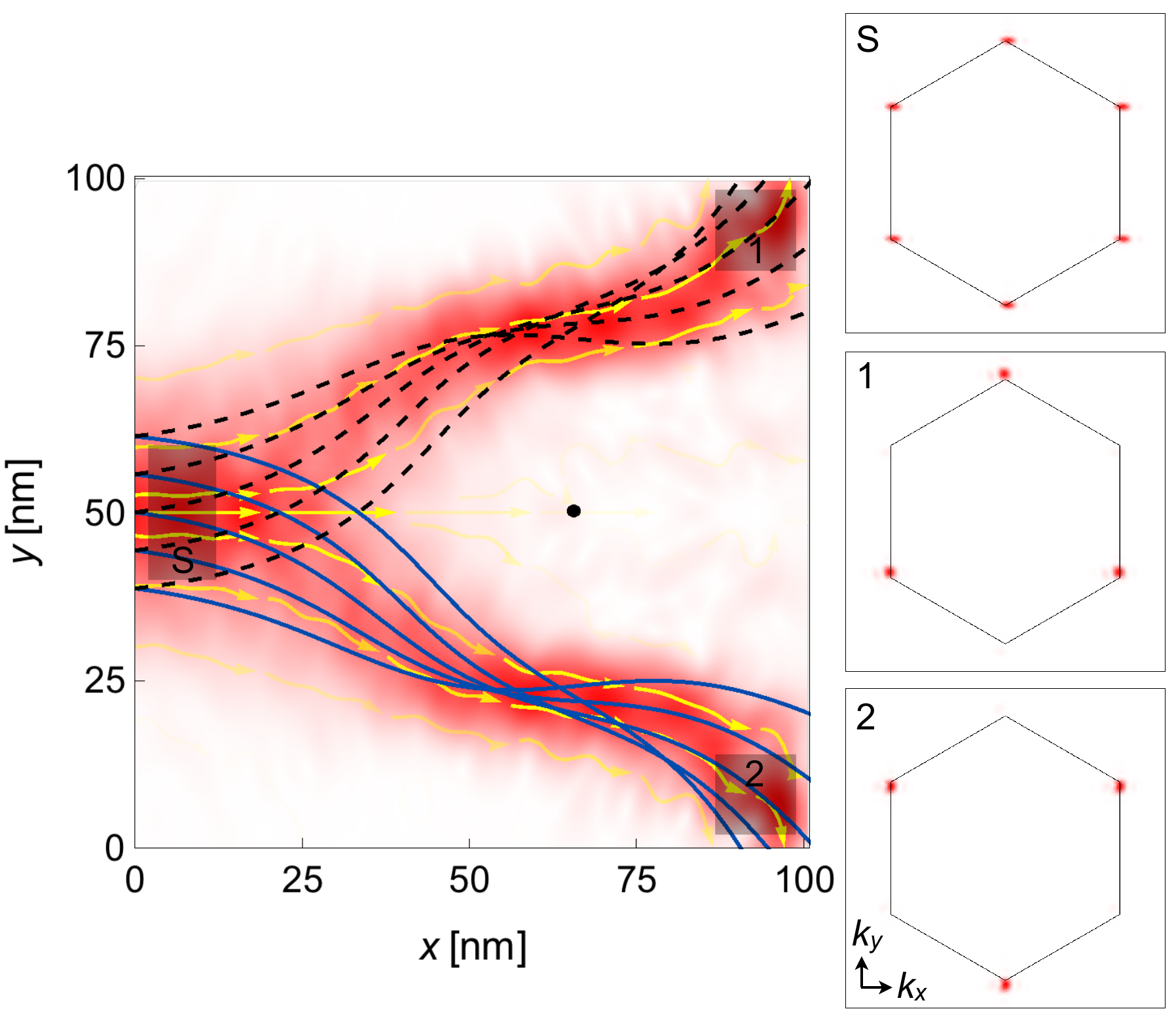}
  \caption{Left: Current flow in the deformed graphene nanoribbon. The center of the deformation is
    indicated by the black dot. The current (red color shading and yellow arrows) is bent around the
    deformation centered at $\v r_c=(0.65L_x, 0.50L_y)$, see the black dot. The current flow lines
    in geometric optics approximation (blue solid and black dashed curves) agree well with the
    current obtained from the NEGF method. Right: The projection $\mc{P}_i(\v k)$ in the gray-shaded
    regions shows that the current is injected unpolarized and split into two beams of valley
    polarized currents.}
  \label{fig:3}
\end{figure}

\subsection{Measurement of the valley polarization}
\label{sec:ValleyPol}

The valley polarization of a state $\ket{\ap}$ can be measured by its projection
$ P(\v k)= \abs{\braket{\psi(\v k) | \ap}}^2$ onto the graphene lattice eigenstates%
\footnote{Here, we chose the eigenstates of the lattice Hamiltonian \eq{1}. Projecting onto the
  eigenstates of the Dirac Hamiltonian \eq{18} gives qualitatively identical results.}
\begin{equation}
  \label{eq:23}
  \psi_n(\v k)= 
  \begin{cases}
     e^{\I \v k \v r_n} & n \in A,\\
     \sg e^{\I \v k (\v r_n - \v \dl)} \frac{f(\v k)}{\Abs{f(\v k)}} & n \in B,
  \end{cases}
\end{equation}
where
\begin{equation}
  \label{eq:22}
  f(\v k)= - t e^{-3\I k_x d_0}\bigl[1+2e^{3 \I k_x d_0/2}\cos(\sqrt{3} k_yd_0/2)\bigr],
\end{equation}
$\sg= \text{sign}(E)$ and $\v \dl= (-1,\sqrt{3}) d_0/2$ is the vector that connects the atoms in
sublattice A with the atoms in sublattice B. This projection represents the occupied states in $k$
space and hence allows us to determine the valley polarization.  Within the NEGF formalism, it can
be transformed to
\begin{equation}
  \label{eq:15}
  P_i(\v k)= \braket{\psi(\v k) | G\, {\Gm^\text{in}_S}\, G^\dg | \psi(\v k)}_{\mc{R}_i},
\end{equation}
where the projection can be calculated over a finite region $\mc{R}_i$ of the system, see for
example the gray-shaded regions in \fig{2}. The spectral density $P_i(\v k)$ is integrated in
hexagonal regions $\mc{K}^\pm$ around the valleys $K^\pm$,
\begin{equation}
  \label{eq:16}
  \mc{P}_i^\pm= \int_{\v k \in \mc{K}^\pm} d^2k \, P_i(\v k),
\end{equation}
see for example the small hexagons in \fig{2} (right). The valley polarization is then given by
\begin{equation}
  \label{eq:17}
  \mc{P}_i= \frac{\mc{P}_i^+-\mc{P}_i^-}{\mc{P}_i^++\mc{P}_i^-}.
\end{equation}
For $\mc{P}_i=\pm 1$ the electrons are localized exclusively at the $K^{\pm}$ valleys and hence, are
completely valley polarized.  However, this relative measure can be misleading because it is
independent of the current density and hence does not assess how much polarized current is flowing
in the system.  Our aim is to propose an efficient device, where a high valley polarization of the
electrons comes along with a high transmission of the injected current. Hence, we assess the
efficiency of our device by multiplying the valley polarization with the corresponding (normalized)
transmission,
\begin{equation}
  \label{eq:eff}
  \mc{Q}_i=\mc{P}_i \, T_{iS}/T_{\text{tot}}.
\end{equation}

\section{Results}
\label{sec:Results}

\subsection{Current flow in flat and deformed graphene nanoribbons}

\begin{figure}[t]
  \centering
  \vspace*{0.5mm}
  \includegraphics[scale=0.285]{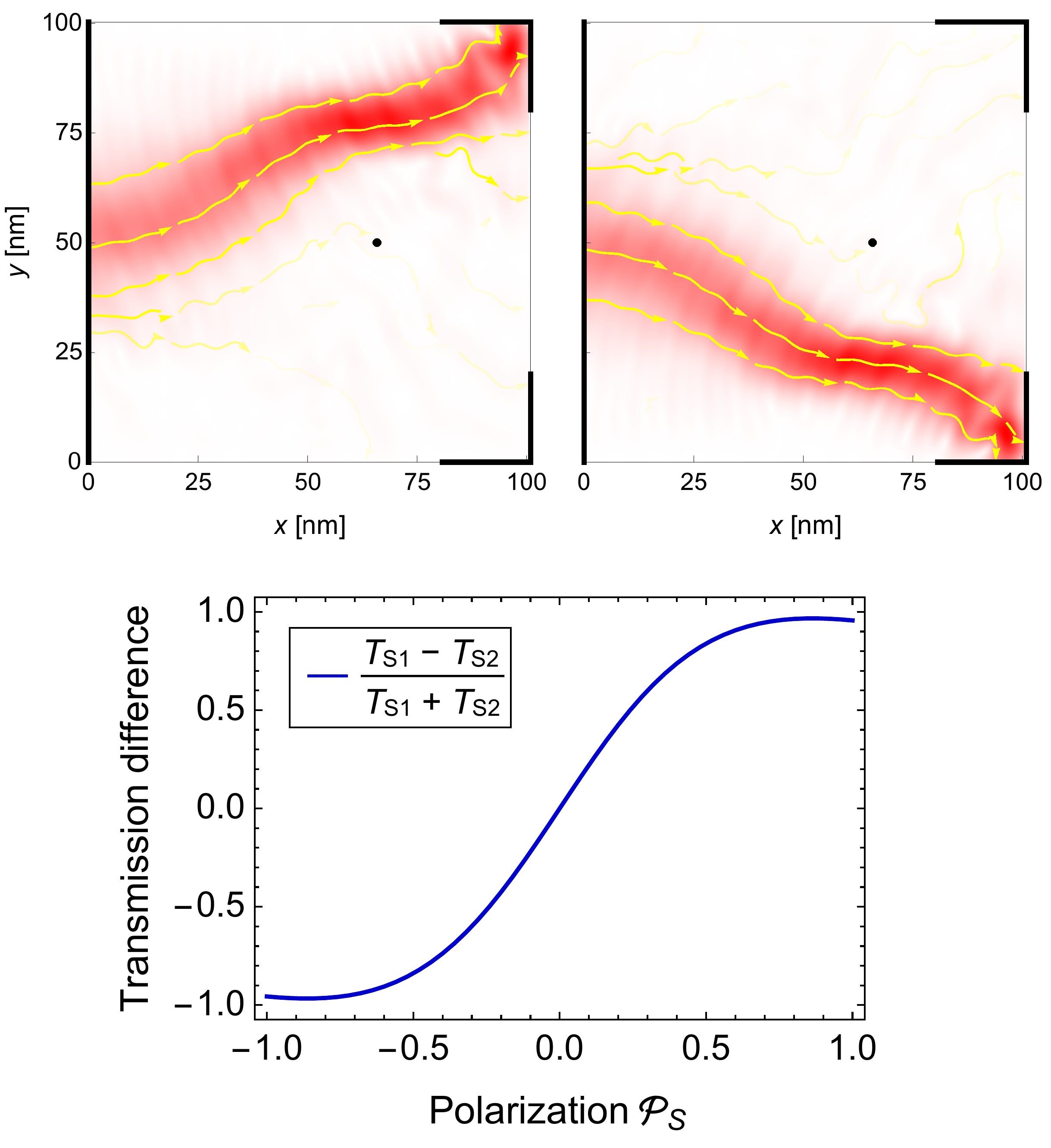}
  \caption{Current flow in a deformed graphene nanoribbon. Top: When the injected current is
    polarized in the $K^+$ valley it is bent upwards (left), while in the $K^-$ valley it is bent
    downwards (right). Bottom: The (relative) difference of the transmission between the source and
    the contacts 1 and 2 as a function of the polarization of the injected current.}
  \label{fig:4}
\end{figure}

The current flow in a flat graphene nanoribbon without any deformation is shown in
\fig{2}. Electrons with energy $E=225 \un{meV}$ are injected at the left ribbon edge. The current
flows straight along the system indicating ballistic electron transport. Such current paths in
graphene nanoribbons have been observed experimentally very recently \cite{Tetienne2017}. The
projection $\mc{P}_i(\v k)$ shows occupied states in all six valleys.%
\footnote{Only two of these valleys are inequivalent, but for clarity we prefer to draw all six of
  them. The numerical exact agreement of the projections in equivalent valleys provides also a
  consistency check of our computer code.}  It confirms that the current is injected unpolarized and
remains unpolarized after traversing the system.

The nanoribbon is then deformed in the way described by \eq{2} with width $r_0=0.55 L_x$, height
$h_0=0.22r_0$ and center $\v r_c=(0.65L_x, 0.50L_y)$. This deformation induces a strain $\dl_{nm}$
of max $1.2\%$, which corresponds to changes of the coupling matrix elements $t_{ij}$ by max $4
\%$. Note that we assume that the height and width of the deformation are proportional, which seems
to be natural for a deformation caused by the tip of a microscope. It can be observed that due to
the interaction with the curvature and the pseudo-magnetic field the current is bent around the
deformation, see \fig{3}. The calculation of $\mc{P}_i(\v k)$ in the gray-shaded regions indicates
clearly that almost all electrons in the upper beam occupy the $K^+$ valley while nearly all
electrons in the lower beam are in the $K^-$ valley. This follows from the fact that the
pseudo-magnetic field, sketched in \fig{1}, acts with opposite signs on the electrons in different
valleys and hence, separates them spatially. Moreover, its interpolating form has the ability to
focus the electron beams. The different valley polarizations of the two beams give rise to a finite
valley voltage \cite{Settnes2017} between the contacts 1 and 2 (the real voltage is zero as the
total currents in both beams are equal).

The current flow lines in the geometric optics approximation, which are indicated by the solid black
curves in \fig{2} and \fig{3}, agree qualitatively with the current obtained by the NEGF
method. Hence, the current flow lines can be used for a fast estimation of the local current flow,
although they do not provide quantitative information on the current density.

In \fig{4} (top), valley polarized currents are injected at the left ribbon edge. The current
flow paths confirm that the effect of the deformation depends strongly on the valley spin, because
electrons at the $K^+$ valley are bent upwards while electrons at the $K^-$ valley are bent
downwards. Additionally, we calculate the transmission $T_{1S}$ and $T_{2S}$ between the source at
the left and the contacts at the right (see the thick black bars at the edges of \fig{4}
(top)). Their (relative) difference $\frac{T_{S1}-T_{S2}}{T_{S1}+T_{S2}}$ is depicted in \fig{4}
(bottom) and demonstrates that the proposed setup can be used to measure the valley
polarization by means of the current detected at the contacts 1 and 2.

\begin{figure}[t]
  \centering
  \includegraphics[scale=0.35]{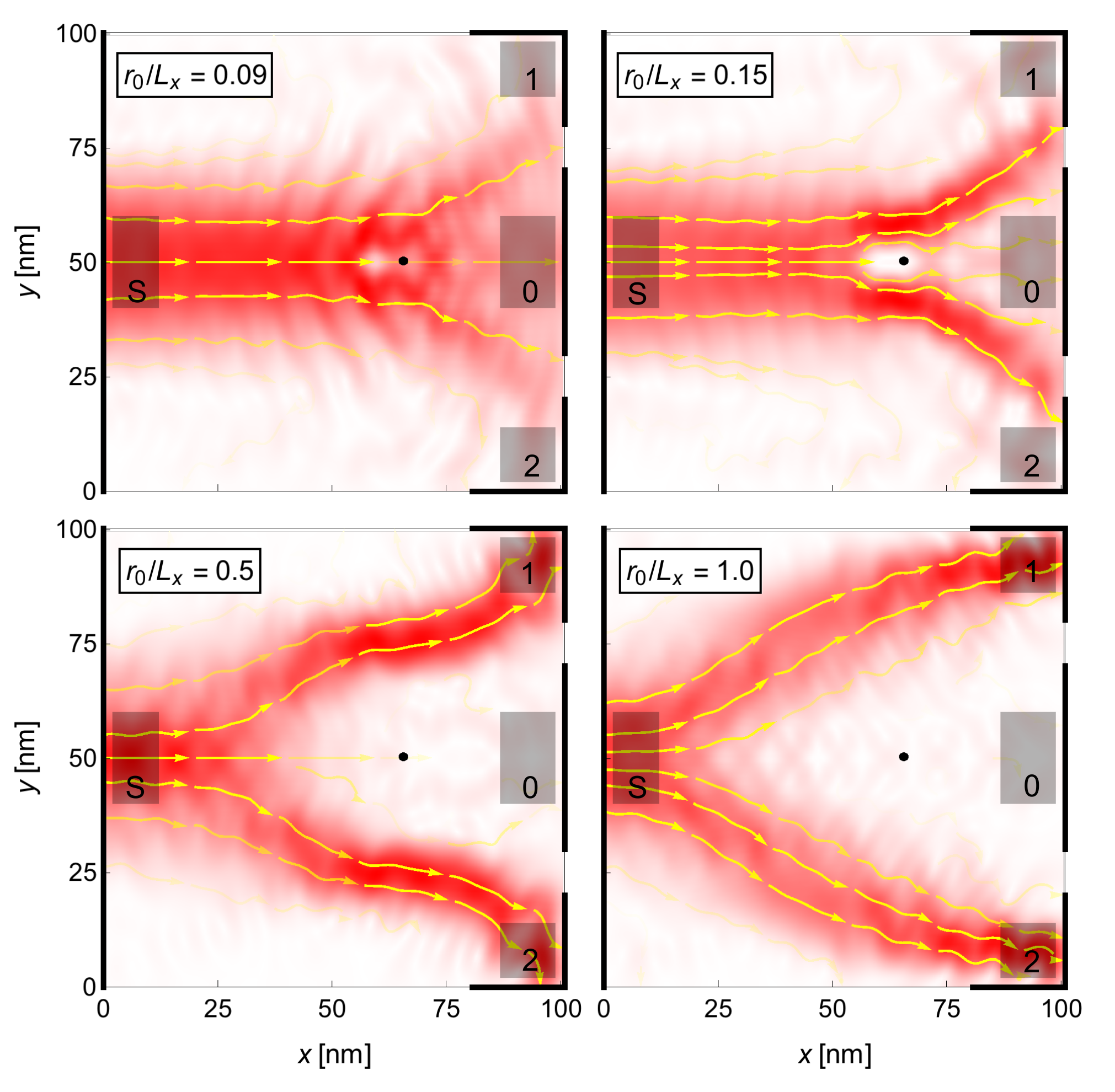}
  \caption{Current flow in a graphene nanoribbon with deformations of various width $r_0$ (see the
    insets) and height $h_0=0.22r_0$. The deflection of the current increases with increasing
    deformation.}
  \label{fig:5}
\end{figure}

\begin{figure*}[t]
  \centering
  \includegraphics[scale=0.55]{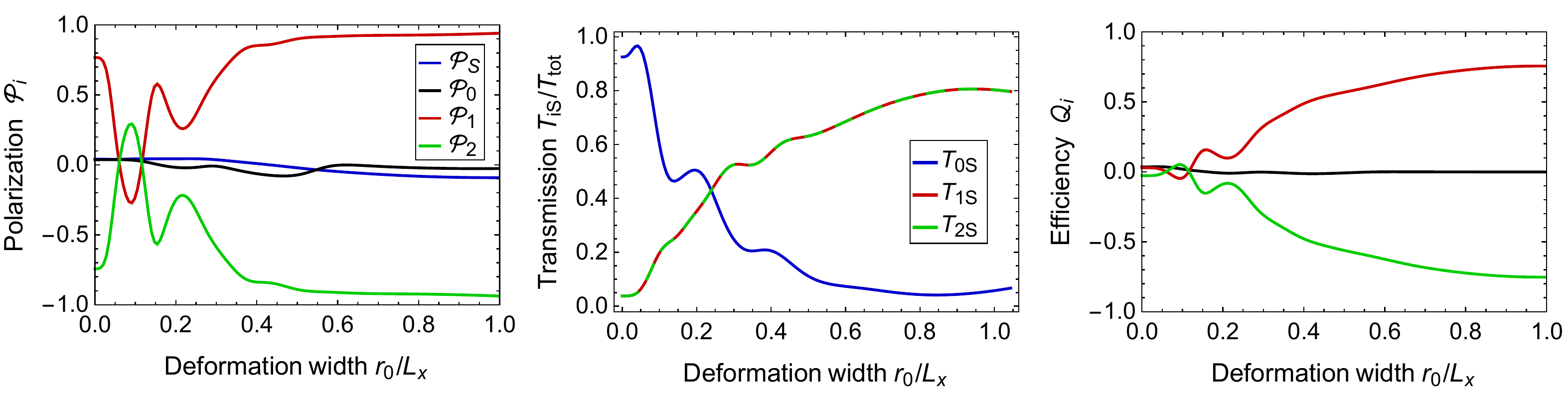}
  \caption{Valley polarization $\mc{P}_i$, transmission $T_{iS}$ and efficiency $\mc{Q}_i$ as a
    function of the deformation width $r_0$ and height $h_0=0.22r_0$. For widths $r_0>0.5L_x$ a
    highly polarized current is obtained which comes along with a high transmission making the
    device very efficient.}
  \label{fig:6}
\end{figure*}

\subsection{Valley polarization as a function of deformation}

In the following, we study in more detail how the valley polarization is affected by deformations
described by \eq{2}. We consider the variable width $r_0$ and the height $h_0=0.22r_0$, which grows
proportionally to the width. The deformation is centered at $\v r_c=(0.65L_x, 0.50L_y)$. The current
flow patterns for various deformations are depicted in \fig{5} and confirm the expectation that the
current deflection increases with the deformation strength. Additionally, we calculate the valley
polarization $\mc{P}_i$ in the gray shaded regions and the transmission $T_{iS}$ between the source
at the left and the three contacts at the right (see the thick black bars at the edges of
\fig{5}). This allows us to calculate the device efficiency $\mc{Q}_i$. These quantities are shown
as a function of the deformation size in \fig{6}. Note that for symmetry reasons
$\mc{P}_1=-\mc{P}_2$.

 \begin{figure}[t]
   \centering
   \includegraphics[scale=0.62]{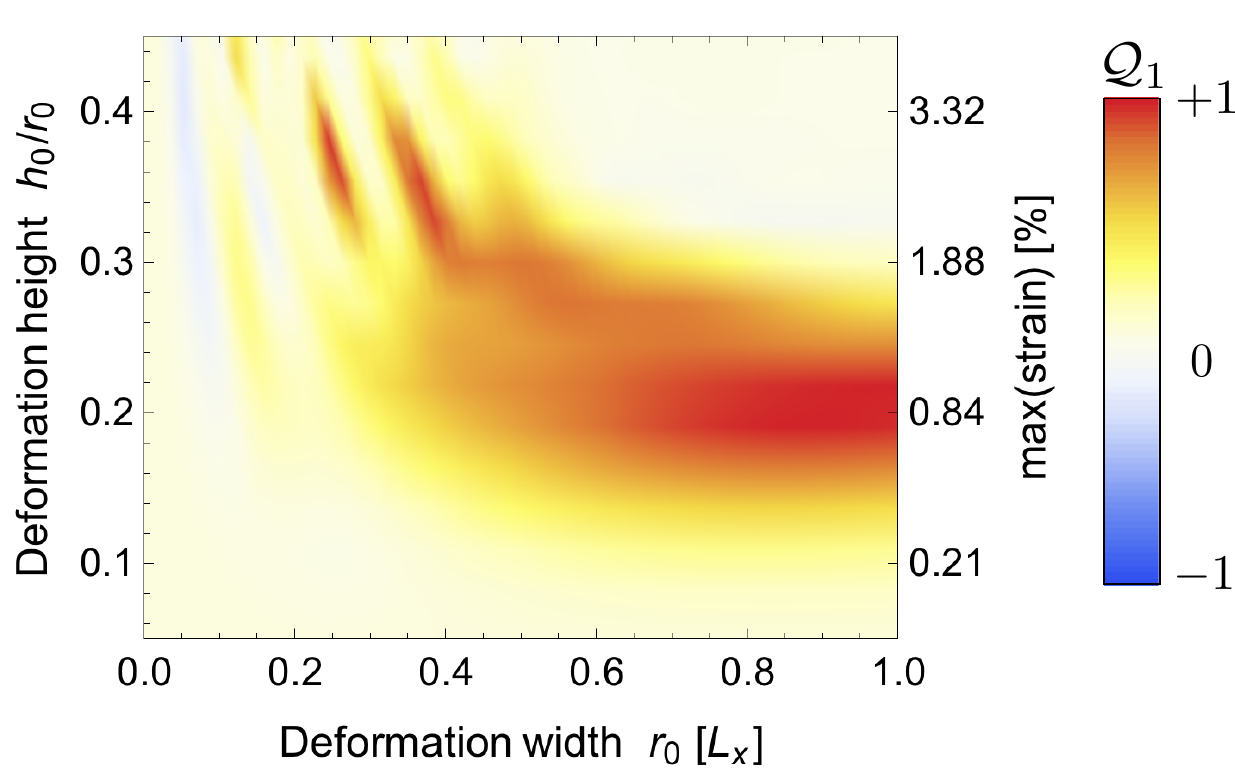}
   \caption{Device efficiency $\mc{Q}_1$ (color density) calculated at contact 1 as a function of
     the deformation width $r_0$ and height $h_0$. Highly efficient valley polarization of the
     electrons is observed for a broad set of parameters. The optimal deformation heights
     ($h_0 \sim 0.2 r_0$) correspond to a maximal strain of $0.9\%$.}
  \label{fig:7}
\end{figure}

 \begin{figure}[t]
   \centering
   \includegraphics[scale=0.58]{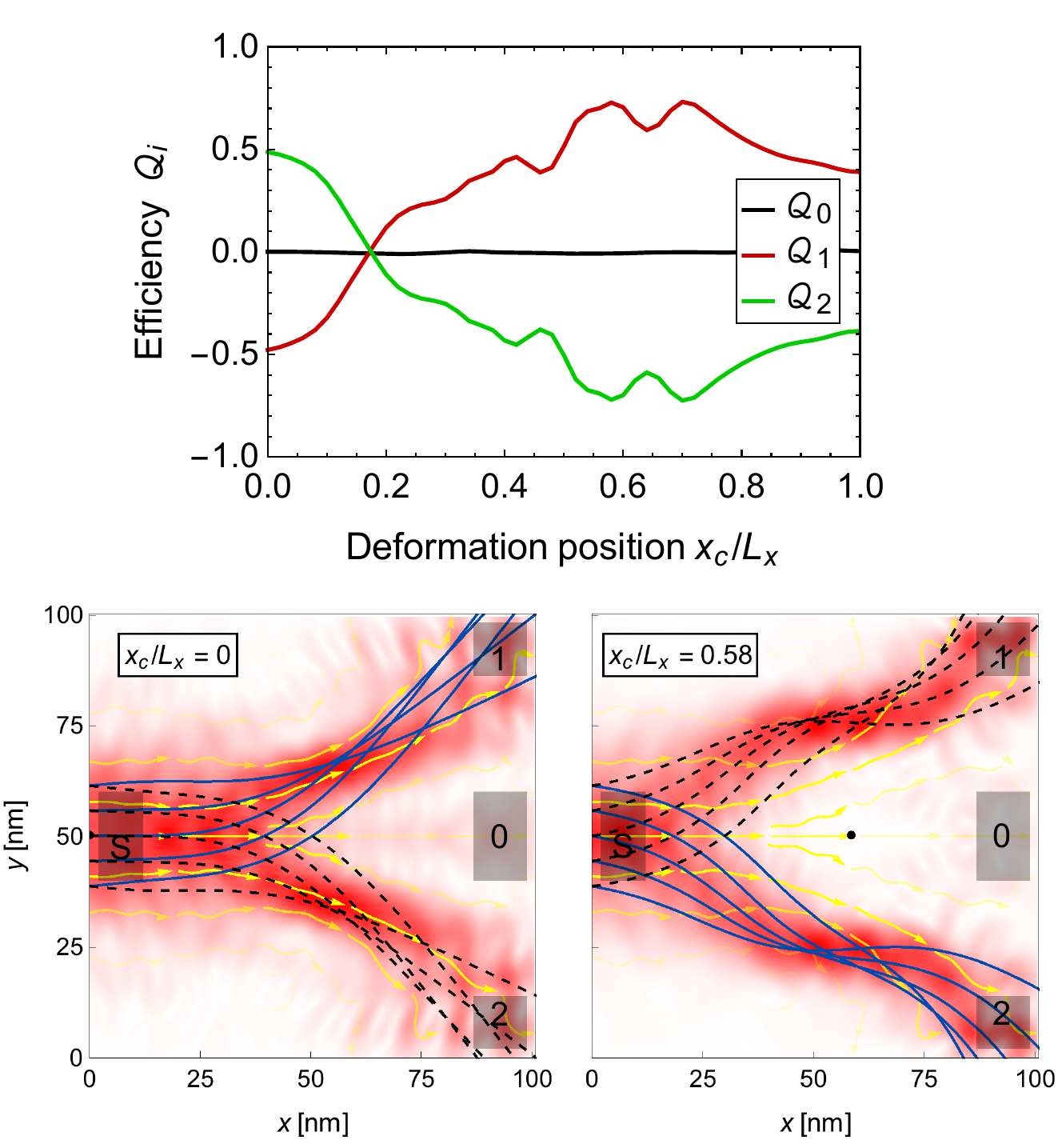}
   \caption{Device efficiency as a function of the position $x_c$ of a deformation with width
     $r_0=0.55L_x$ and height $h_0=0.22r_0$. The efficiency approaches its maximum for
     $x_c \approx 0.6$. For $x_c=0$ the sign of the polarization is reversed because the electrons
     are injected into the center of the deformation and deflected differently by the
     pseudo-magnetic field (compare the solid black and blue trajectories in the local current
     flow).}
  \label{fig:8}
\end{figure}

In the regions S and 0 the electrons are unpolarized $\mc{P}_{S/0} \approx 0$ and their polarization
is independent of the deformation. In the regions 1 and 2 highly polarized electrons
$\mc{P}_{1/2} \approx \pm 1$ are obtained for $r_0>0.4 L_x$ but, surprisingly, also in an almost
flat system. This behavior can be explained by the fact that the polarization is a relative measure
and does not assess how many electrons take part in the flow. In fact, \fig{2} confirms that in a
flat nanoribbon only very few electrons are transfered to the right corners. Their polarization may
stem from the trigonal warping of the Dirac cones, see \sect{TriWarp} for the discussion. Moreover,
for $r_0 =0.09 L_x$ we can observe that the sign of the polarization in the regions 1 and 2 is
inverted. This behavior may originate from reflections at the ribbon edges as the deformation is not
sufficiently strong to deflect a substantial part of the current, see \fig{5}.

\Fig{6} (middle) shows that the transmissions $T_{1S}$ and $T_{2S}$ increase with the deformation
while $T_{0S}$ decreases. This quantifies our observation from \fig{5} that the deflection of the
injected current grows with the increasing deformation. Note that for symmetry reasons
$T_{1S}= T_{2S}$. The efficiency of the device, shown in \fig{6} (right), also increases with the
deformation size. It can be observed that for $r_0>0.5L_x$ a highly efficient device is obtained
where almost all of the injected electrons are split up into two fully polarized electron beams (see
also the current flow paths in \fig{5}). The deformation needs to be rather wide in order to catch a
significant part of the injected current by the first lobe of the pseudo-magnetic field (see the red
lobe closest to the source in \fig{1}) and to split it into two valley polarized beams.  Moreover,
the functional dependence between the transmission and the deformation size in \fig{6} (middle) can
also be used to determine the deformation of the system from a transport measurement.

Next, we vary both, the width $r_0$ and height $h_0$ of the deformation independently while keeping
its center fixed at $\v r_c=(0.65L_x,0.5L_y)$. The efficiency at the contact 1 is shown in
\fig{7}. Highly efficient valley polarization of the electrons is observed for wide deformations
$0.5<r_0/L_x<1$ with a height $0.15 <h_0/r_0<0.3$, which corresponds to a maximal strain of
$0.6-1.2\%$. Hence, in the suggested device it is favorable that the pseudo-magnetic field extends
over a significant part of the system in order to collect a large part of the injected current. Note
that the efficiencies satisfy $\mc{Q}_2=-\mc{Q}_1$.

Finally, we vary the position of the deformation $\v r_c=(x_c L_x, L_y/2)$ in the system while
keeping its width $r_0=0.55 L_x$ and height $h_0=0.22r_0$ fixed. The device efficiency $\mc{Q}_1$,
shown in \fig{8} (left), approaches its maximum when the deformation is placed slightly behind the
ribbon center, i.e. at $x_c \approx 0.6 L_x$. Surprisingly, we observe that for a deformation, which
is centered at the left ribbon edge ($x_c=0$), the sign of the efficiency and hence, the sign of the
polarization gets reversed. This effect can be explained by the fact that the electrons are injected
in the center of the pseudo-magnetic field which acts differently there, see \fig{1}. The classical
trajectories in \fig{8} (right) clearly indicate that the electrons in the two valleys (solid black
and blue curves) are deflected differently. However, injecting electrons precisely in the center of
a deformation may be experimentally challenging.

\subsection{Valley polarization by the trigonal warping of the Dirac cones}
\label{sec:TriWarp}

\begin{figure}[t]
  \centering
  \includegraphics[scale=0.4]{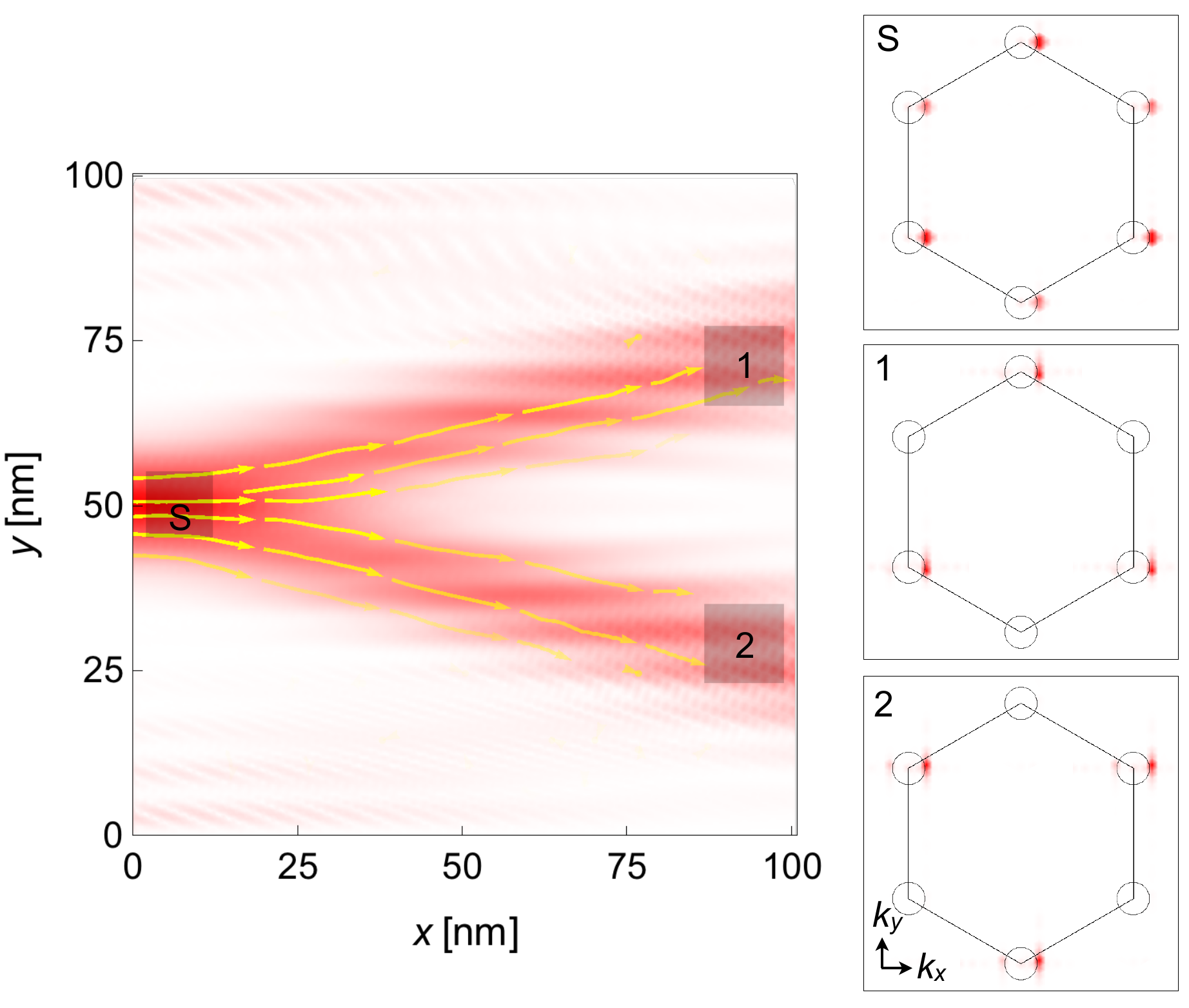}
  \caption{Electron flow at a high energy ($E=1.26\un{eV}$) in a graphene nanoribbon without
    deformation. The current is split into two valley polarized beams due to the trigonal warping of
    the Dirac cones, which sends electrons from different valleys in different directions. The
    occupied states are shifted approximately by $q=0.3/d_0$ (small circles), which is consistent
    with the high injection energy. Note that in this figure the regions around the Dirac point have
    not been magnified.}
  \label{fig:9}
\end{figure}

As a side remark, we comment that another way to generate a valley polarized current might be to
inject electrons at high energy into a flat graphene nanoribbon. In \fig{9} (left), the electrons
that are injected with energy $E=1.26\un{eV}$ are split into two valley polarized beams.  The
splitting of the electron beam as well as the valley polarization are due to the trigonal warping of
the Dirac cones. This means that at high energies the Fermi surface in graphene is no longer round
but becomes triangular, which sends the electrons from the two valleys in two different
directions. Details on the graphene valley polarizer based on the trigonal warping of the Dirac
cones at high energies can be found in Refs.~\cite{Garcia-Pomar2008, Yang2017}. Here, \fig{9}
(right) provides also a consistency check of our calculations because the occupied states are
shifted by $|\v q|= |E|/\hbar v_F= 0.3/d_0$ (small circles) in perfect agreement with the injection
energy and the dispersion relation.

\section{Conclusions}
\label{sec:conclusions}

We studied theoretically the current flow in graphene nanoribbons with smooth out-of-plane
deformations (\fig{1}) by means of a tight-binding model and the NEGF method.  Already for moderate
strains, up to $1\%$, we observed a complete directional splitting of valley currents (\fig{3}) and
a full valley polarization of the separated beams (\fig{3},~\ref{fig:4}). The different
polarizations of the two beams give rise to a finite valley voltage. We studied the influence of
the deformation's strength (\fig{5}), shape (\fig{7}) and position (\fig{8}) on the current
splitting efficiency by measuring the transmission between four contacts (\fig{6}).

To complete the understanding, we established a connection between the current flow paths and
classical trajectories of particles moving in curved space in presence of a magnetic field. We
derived the latter picture in an earlier work from the effective Dirac equation in deformed graphene
\cite{Stegmann2016}.

These model calculations demonstrate the feasibility of the proposal of a deformation sensor
nanodevice.  The small size of the system ($100\times 100\un{nm}$) is only a consequence of our
numerical limitations and we expect the phenomena to persist at larger scales as long as the
transport stays ballistic.  At larger scales ($L>100\un{nm}$) and lower energies
($E<0.1 t_0 \approx 280\un{meV}$) the semiclassical approximation is expected to give even better
current flow predictions.

An interesting open topic is the exact form of the bump created in experiments (strain due to
lattice mismatch with the substrate, microscope tip, or air pressured membrane) and the spatial
distribution of the carbon atoms obtained by taking into account various binding and relaxation
mechanisms.  Another element of the system, which can be developed further, is the model of the
contacts. We plan to quantify the coherence of injected electrons via graphene or hetero-metallic
leads and study their influence on polarization efficiency in the transport.

\begin{acknowledgments}
  TS acknowledges financial support from CONACYT Proyecto Fronteras 952 and from the UNAM-PAPIIT
  research grant IA101618. NS acknowledges the support by the ``Deutsche Forschungsgemeinschaft''
  (DFG) through project B7 of the Collaborative Research Centre (SFB) 1242. TS thanks Reyes Garcia
  for computer technical support. We thank Thomas H. Seligman for useful discussions.
\end{acknowledgments}

\bibliography{gvp}

\end{document}